\documentclass[reprint,amsmath,amssymb,aps,prl,nofootinbib]{revtex4-2}

\usepackage{bm}
\usepackage{graphicx}
\usepackage[colorlinks=true,allcolors=blue]{hyperref}
\usepackage{amsmath,amssymb,booktabs}

\usepackage{tablefootnote}
\usepackage{ragged2e}

\begin{document}

\title{Probing departures from Schwarzschild geometry with the S301 star and kilometre-baseline interferometry}

\author{Rodrigo P. Silva}
\affiliation{CFisUC, Departamento de F\'isica, Universidade de Coimbra, 3004-516 Coimbra, Portugal}
\affiliation{Faculdade de Engenharia, Universidade do Porto, Rua Dr. Roberto Frias, 4200-465 Porto, Portugal}
\affiliation{CENTRA---Centro de Astrof\'isica e Gravita\c{c}\~ao, IST, Universidade de Lisboa, 1049-001 Lisboa, Portugal}

\author{Alexandre C. M. Correia}
\affiliation{CFisUC, Departamento de F\'isica, Universidade de Coimbra, 3004-516 Coimbra, Portugal}
\affiliation{LTE, Observatoire de Paris, Universit\'e PSL, Sorbonne Universit\'e, CNRS, 75014 Paris, France}

\author{Paulo J. V. Garcia}
\affiliation{Faculdade de Engenharia, Universidade do Porto, Rua Dr. Roberto Frias, 4200-465 Porto, Portugal}
\affiliation{CENTRA---Centro de Astrof\'isica e Gravita\c{c}\~ao, IST, Universidade de Lisboa, 1049-001 Lisboa, Portugal}

\date{\today}

\begin{abstract}
	The S301 star reaches a pericentre one tenth that of S2, opening a new regime for probing black-hole spin and departures from Schwarzschild. We develop a post-Newtonian secular Hamiltonian framework for bound geodesics in stationary, axisymmetric spacetimes and compute the maximum projected astrometric deviation over one orbit. At the expected GRAVITY+ precision, S301 can probe several nonrotating geometries and sizeable spin-induced deviations, while kilometre-baseline interferometry can probe much smaller departures, establishing S301 as a powerful test of the spacetime geometry of Sgr A*.
\end{abstract}

\maketitle

\emph{\textbf{Introduction.---}}
The centre of the Galaxy harbours the \(4.3\times10^6\,M_\odot\) compact object Sgr~A* \citep{2008ApJ...689.1044G,2009ApJ...692.1075G}. The relativistic orbit of S2 has already revealed the gravitational redshift and Schwarzschild precession \citep{2018A&A...615L..15G,2019Sci...365..664D,2020A&A...636L...5G,2021A&A...647A..59G,2022A&A...657L..12G}. These measurements establish leading relativistic effects in the weak-field regime, but do not uniquely determine the spacetime geometry, since many non-Schwarzschild metrics share the same leading Schwarzschild behaviour. Moreover, at current sensitivity, measurements of S2 are not sensitive to the spin of Sgr~A* \citep{2024A&A...692A.242G}. Forecasts indicate that detecting frame dragging requires stars on substantially more compact orbits than S2 \citep{2018MNRAS.476.3600W,2023MNRAS.522.5828C}.

Stellar orbits around Sgr~A* also probe the distribution of matter and possible departures from general relativity (GR) in the Galactic Centre \citep{2023PhRvD.108j1303D,2023MNRAS.524.1075F,2019MNRAS.489.4606G,2023RPPh...86j4901D}. Although these orbits remain in the weak-field regime, their secular evolution can be sensitive to higher-order corrections that are strongly suppressed for less compact trajectories. A useful sensitivity forecast must therefore establish not only whether an alternative spacetime modifies the orbit, but whether the resulting deviation is distinguishable at a given astrometric precision. Apsidal precession is particularly important because it provides the dominant secular signal, while nodal precession carries complementary information about the spin orientation and can help distinguish rotational effects from spherically symmetric deformations.

The recently discovered star S301 provides such an opportunity \citep{2026Natur.657..359A}. Its pericentre lies at about \(140\) Schwarzschild radii, compared with about \(1400\) for S2. Since frame dragging enters at higher post-Newtonian (PN) order than Schwarzschild precession and grows rapidly for more compact orbits, S301 is substantially more sensitive than S2 to the spin of Sgr~A*. More generally, its compact orbit opens a regime in which small departures from the Schwarzschild geometry can accumulate into measurable secular displacements. Previous studies of S301 have primarily emphasised its sensitivity to the Kerr spin of Sgr~A* \citep[e.g.][]{2026Natur.657..359A,2026arXiv260724931P}. Here, we instead ask how well its orbit can distinguish Schwarzschild from a broader class of stationary, axisymmetric spacetimes.

To address this question, we develop a unified Hamiltonian framework for computing apsidal and nodal precessions from a common secular Hamiltonian, rather than deriving each case separately from the geodesic equations. The formalism makes explicit how the weak-field coefficients of a metric enter the secular observables and can be systematically extended to other stationary, axisymmetric spacetimes and higher PN orders. We then compute the projected astrometric residuals between the Schwarzschild prediction and the alternative-spacetime trajectories considered by \citet{2022ApJ...930L..17E}, together with the Simpson--Visser geometry \citep{2019JCAP...02..042S}, over one orbit of S301. By comparing the maximum residual with the expected astrometric precision of GRAVITY+ and that envisaged for a prospective kilometre-baseline interferometer, we determine which departures from the Schwarzschild metric can be probed with stellar astrometry around Sgr A*.

\emph{\textbf{Secular framework.---}}
We consider stationary, axisymmetric spacetimes admitting a weak-field PN expansion. Writing the metric in \(3+1\) form \citep[e.g.][]{2007gr.qc.....3035G, 2010grav.book.....P},
\begin{equation}
	ds^2=-\alpha^2c^2dt^2
	+\gamma_{ij}(dq^i+\beta^ic\,dt)(dq^j+\beta^jc\,dt),
	\label{eq:31metric}
\end{equation}
where \(c\) is the speed of light, \(\alpha\) is the lapse function, \(\beta^i\) is the shift vector, \(\gamma_{ij}\) is the spatial metric, \(t\) is the time coordinate, and \(q^i\) are the spatial coordinates, with \(i,j=1,2,3\).
The Hamiltonian generating coordinate-time evolution for a particle of mass \(m\) is \citep[e.g.][]{2007gr.qc.....3035G, 2018ApJS..237....6B}
\begin{equation}
	\mathcal H
	=\alpha c\sqrt{m^2c^2+\gamma^{ij}p_ip_j}
	-c\beta^ip_i,
	\label{eq:exactH}
\end{equation}
where \(p_i\) are the momenta conjugate to \(q^i\).

Expanding the \(3+1\) fields in the weak-field regime and subtracting the constant rest-mass contribution, we introduce a formal dimensionless expansion parameter \(\epsilon \sim v/c\), used to keep track of the post-Newtonian ordering, and write
\begin{equation}
	\mathcal H=\mathcal H_0+\epsilon^2\mathcal H_1
	+\epsilon^3\mathcal H_{1.5}
	+\epsilon^4\mathcal H_2+\mathcal O(\epsilon^5),
	\label{eq:pnH}
\end{equation}
where \(\mathcal H_0=p^2/(2m)-m\mu/r\) is the Kepler Hamiltonian, \(\mu=\mathcal G M_\bullet\) is the gravitational parameter of Sgr A*, \(\mathcal G\) is the gravitational constant, \(M_\bullet\) is the mass of Sgr A*, and \(p^2\equiv h^{ij}p_i p_j\), with \(h^{ij}\) denoting the inverse Euclidean metric in the chosen coordinate system. The term \(\mathcal H_{1.5}\) contains the leading coupling between the black-hole spin and the orbital angular momentum, while the PN coefficients of the lapse, shift, and inverse spatial metric determine the perturbing Hamiltonians at each order.

In Eq.~(\ref{eq:pnH}), we have a perturbation to an integrable problem. We therefore seek a canonical transformation that shifts the perturbation to higher order, yielding a new Hamiltonian that is integrable to the desired perturbative order. We achieve this by means of a near-identity Lie transformation \citep[cf.][]{1966PASJ...18..287H, 1969CeMec...1...12D, 1983rsm..book.....L, 1988CeMec..43..193R, 2013MNRAS.428.2295B}. 
A useful set of canonical variables for describing the transformed Hamiltonian is provided by the Delaunay variables \citep[e.g.][]{2023dyps.book.....T},
\begin{equation}
	L=m\sqrt{\mu a},\qquad
	G=L\sqrt{1-e^2},\qquad
	H=G\cos I,
\end{equation}
where \(m\) is the mass of the celestial object orbiting Sgr~A*, \(a\) is the semi-major axis, \(e\) is the eccentricity, and \(I\) is the inclination with respect to a chosen reference plane. Their conjugate angles are the mean anomaly \(\ell\), the argument of pericentre \(\omega\), and the longitude of the ascending node \(\Omega\).

The resulting Hamiltonian is 
\begin{align}
	\langle\mathcal H^\prime\rangle
	=
	{}&
	\mathcal H_0
	+\langle\epsilon^2\mathcal H_1\rangle
	+\langle\epsilon^3\mathcal H_{1.5}\rangle \nonumber\\
	&
	+\langle \epsilon^4\mathcal H_2\rangle
	+\frac{1}{2}
	\left\langle\epsilon^4
	\{\widetilde{\mathcal H}_1,S_1\}
	\right\rangle
	+\mathcal O(\epsilon^5),
	\label{eq:secularH}
\end{align}
where \(\langle\cdots\rangle\) denotes the orbital average,
\begin{equation}
	\langle F\rangle
	=
	\frac{1}{2\pi}
	\int_0^{2\pi}
	F(L, G, H,\ell,\omega,\Omega)\,d\ell.
\end{equation}
Here, \(\{\cdot,\cdot\}\) denotes the Poisson bracket, \(S_1\) is the first-order generating function of the canonical transformation, and \(\widetilde{\mathcal H}_1\) denotes the oscillatory part of the first-order perturbation, defined as
\begin{equation}
	\widetilde{\mathcal H}_1
	=
	\mathcal H_1-\langle\mathcal H_1\rangle.
\end{equation}

Using Hamilton's equations, we obtain the accumulated precessions over one orbit \citep[e.g.][]{2018LRR....21....7S},
\begin{equation}
	\Delta\omega
	=
	\frac{2\pi}{n}
	\frac{\partial\langle\mathcal H^\prime\rangle}{\partial G},
	\qquad
	\Delta\Omega
	=
	\frac{2\pi}{n}
	\frac{\partial\langle\mathcal H^\prime\rangle}{\partial H},
	\label{eq:precessions}
\end{equation}
where
\begin{equation}
	n \equiv \frac{\partial\langle\mathcal H^\prime\rangle}{\partial L}.
	\label{eq:mean_motion_general}
\end{equation}

As consistency checks, this secular formalism reproduces the leading Schwarzschild apsidal advance,
\begin{equation}
	\Delta\omega_{\rm S}
	=\frac{6\pi\mu}{a(1-e^2)c^2},
	\label{eq:schwarzschildcheck}
\end{equation}
and the leading Kerr frame-dragging contributions,
\begin{equation}
	\begin{aligned}
		\Delta\Omega_{\rm LT}
		&=\frac{4\pi\mathcal GJ_\bullet}
		{c^2\sqrt{\mu a^3}(1-e^2)^{3/2}},
		\\
		\Delta\omega_{\rm LT}
		&=-3\cos I\,\Delta\Omega_{\rm LT},
	\end{aligned}
	\label{eq:kerrcheck}
\end{equation}
where \(J_\bullet\) is the spin of Sgr A* and \(I\) is the inclination relative to the black-hole spin axis.

The derivation of Eqs.~\eqref{eq:exactH}, \eqref{eq:pnH} and \eqref{eq:secularH}, the metric-specific weak-field coefficients, and validations for the Schwarzschild, Kerr, and Simpson--Visser spacetimes are provided in the Supplemental Material.

\begin{figure*}
	\centering
	\includegraphics[width=\textwidth]{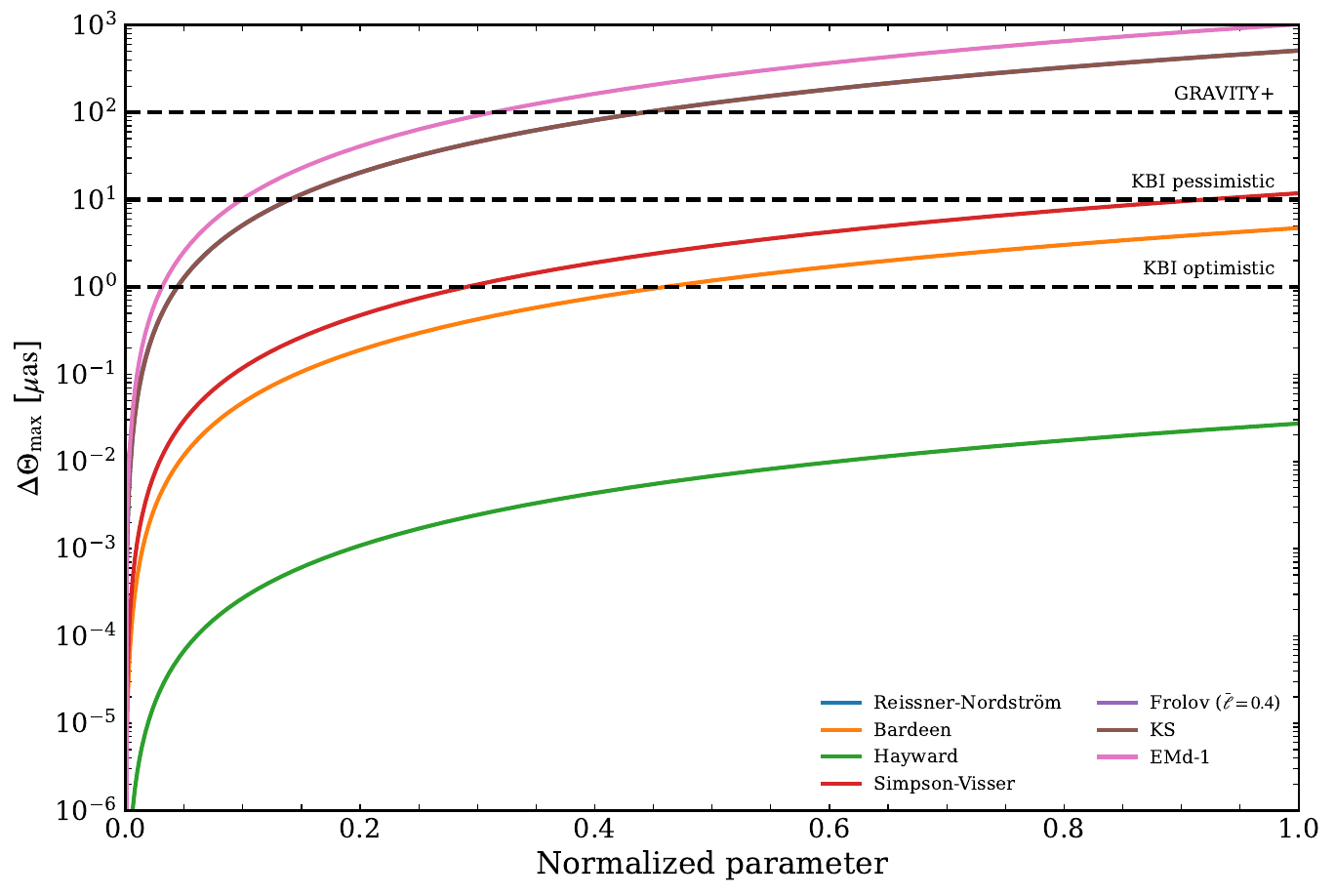}
	\caption{Maximum astrometric residual between one orbit of S301 in a Schwarzschild background and one orbit of S301 in Reissner--Nordstr\"om [blue], Bardeen [orange], Hayward [green], Simpson--Visser [red], Frolov with \(\bar{\ell}=0.4\) [purple], Kazakov--Solodukhin [maroon], and Einstein--Maxwell--dilaton--axion [pink] backgrounds, as a function of their respective normalised parameters. 
	The black dashed lines indicate the expected GRAVITY+ precision and the pessimistic and optimistic KBI precisions.}
	\label{fig:deviations_non_rotating}
\end{figure*}

\begin{table*}[t]
	\centering
	\caption{Comparison of forecast sensitivity limits for nonrotating spacetimes at GRAVITY+ and kilometre-baseline interferometer (KBI) precisions, together with constraints obtained with the Event Horizon Telescope (EHT). Dashes indicate spacetimes that remain unconstrained at the corresponding sensitivities.}
	\label{tab:constraints_no_rotating}
	
	\begin{tabular*}{\textwidth}{@{\extracolsep{\fill}}lccccc@{}}
		\hline
		\textbf{Spacetime}
		& \textbf{Parameter range}
		& \textbf{GRAVITY+ ($100\,\mu\mathrm{as}$)}
		& \textbf{KBI ($10\,\mu\mathrm{as}$)}
		& \textbf{KBI ($1\,\mu\mathrm{as}$)}
		& \textbf{EHT} \citep{2022ApJ...930L..17E} \\
		\hline
		
		Reissner--Nordstr\"om
		& \(0 < \bar{q} \leq 1\)
		& \(0 < \bar{q} \leq 0.44\)
		& \(0 < \bar{q} \leq 0.14\)
		& \(0 < \bar{q} \leq 0.044\)
		& \(0 < \bar{q} \leq 0.84\) \\
		
		Bardeen
		& \(0 < \bar{g} < \sqrt{16/27}\)
		& ---
		& ---
		& \(0 < \bar{g} \leq 0.35\)
		& --- \\
		
		Hayward
		& \(0 < \bar{l} < \sqrt{16/27}\)
		& ---
		& ---
		& ---
		& --- \\
		
		Simpson--Visser
		& \(0 < \bar{\lambda} < 3\)
		& ---
		& \(0 < \bar{\lambda} < 2.8\)
		& \(0 < \bar{\lambda} < 0.87\)
		& --- \\
		
		Frolov (\(\bar{l}=0.4\))
		& \(0 < \bar{q} \leq 0.79\)
		& \(0 < \bar{q} \leq 0.44\)
		& \(0 < \bar{q} \leq 0.14\)
		& \(0 < \bar{q} \leq 0.044\)
		& --- \\
		
		KS
		& \(0 < \bar{\alpha}\)\footnotemark[1]
		& \(0 < \bar{\alpha} \leq 0.63\)
		& \(0 < \bar{\alpha} \leq 0.20\)
		& \(0 < \bar{\alpha} \leq 0.063\)
		& \(0 < \bar{\alpha} \leq 0.79\) \\
		
		EMd-1
		& \(0 < \bar{q} < \sqrt{2}\)
		& \(0 < \bar{q} \leq 0.44\)
		& \(0 < \bar{q} \leq 0.14\)
		& \(0 < \bar{q} \leq 0.044\)
		& \(0 < \bar{q} < 0.87\) \\
		
		\hline
	\end{tabular*}
	
	\footnotetext[1]{The Kazakov--Solodukhin spacetime does not impose an upper limit on the range of its theory parameter \citep{1994NuPhB.429..153K}. Following \citet{2022ApJ...930L..17E}, we restrict the parameter range to \(0 < \bar{\alpha} \leq \sqrt{2}\).}
\end{table*}

\begin{figure*}
	\centering
	\includegraphics[width=\textwidth]{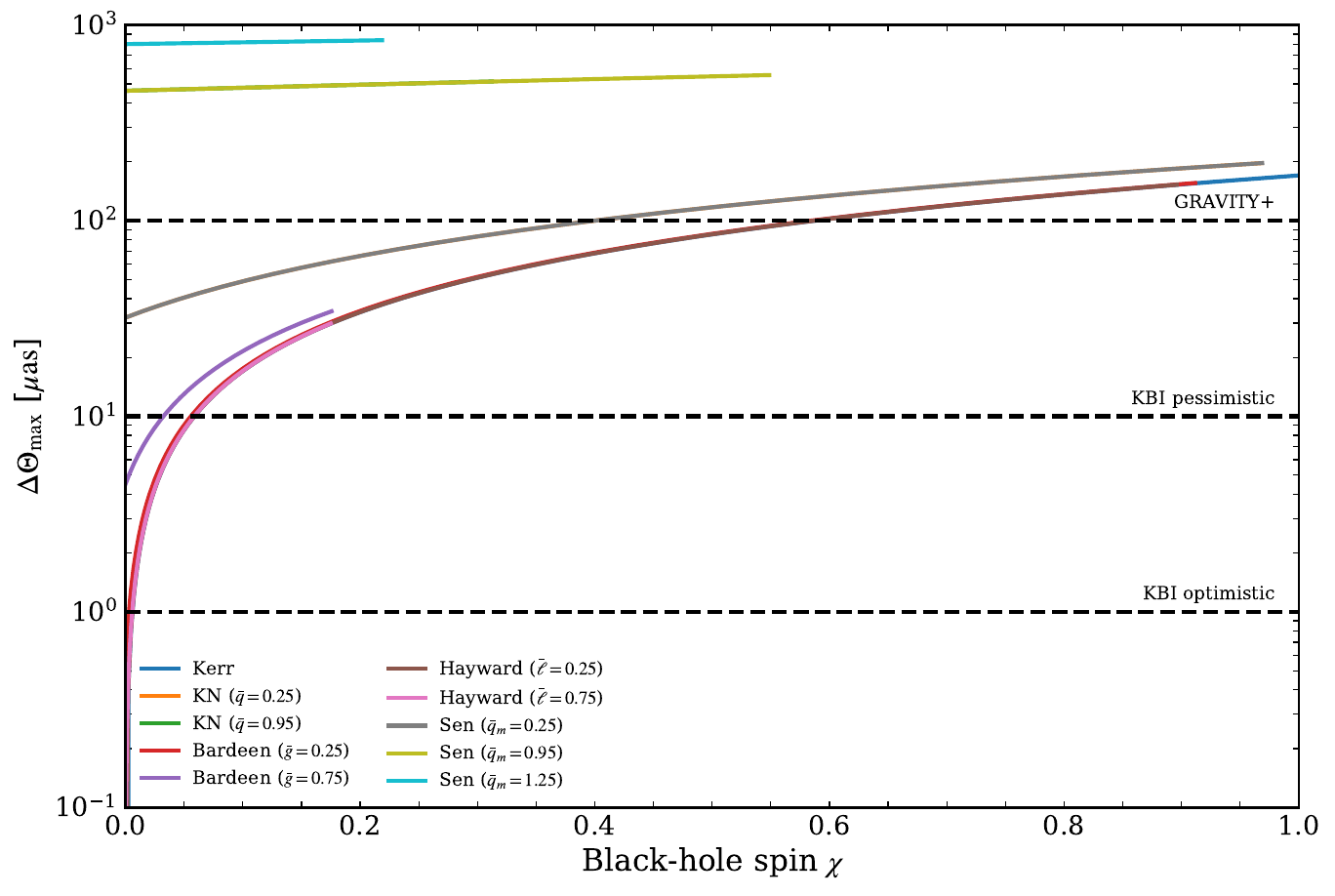}
	\caption{Maximum astrometric residual between one orbit of S301 in a Schwarzschild background and one orbit of S301 in Kerr [blue], Kerr--Newman with \(\bar{q}=0.25\) [orange], Kerr--Newman with \(\bar{q}=0.95\) [green], Bardeen with \(\bar{g}=0.25\) [red], Bardeen with \(\bar{g}=0.75\) [purple], Hayward with \(\bar{\ell}=0.25\) [maroon], Hayward with \(\bar{\ell}=0.75\) [pink], (Kerr)--Sen with \(\bar{q}_{\rm m}=0.25\) [grey], (Kerr)--Sen with \(\bar{q}_{\rm m}=0.95\) [olive], and (Kerr)--Sen with \(\bar{q}_{\rm m}=1.25\) [cyan] backgrounds, as a function of the dimensionless black-hole spin, \(\chi\). 
	The black dashed lines indicate the expected GRAVITY+ precision and the pessimistic and optimistic KBI precisions.
	To maximise the spin effect, we assume that the spin vector is parallel to the orbital angular momentum vector. 
	The extremality condition sets the maximum black-hole spin for each metric, causing some curves to terminate abruptly.}
	\label{fig:deviations_rotating}
\end{figure*}

\begin{table*}[t]
	\centering
	\caption{Comparison of forecast sensitivity limits for rotating spacetimes at GRAVITY+ and KBI precisions, together with constraints obtained with the EHT. Dashes indicate spacetimes that remain unconstrained at the corresponding sensitivities, while crosses indicate cases whose predicted residual exceeds the adopted threshold throughout the allowed spin range.}
	\label{tab:constraints_rotating}
	
	\begin{tabular*}{\textwidth}{@{\extracolsep{\fill}}lccccc@{}}
		\hline
		\textbf{Spacetime}
		& \textbf{Spin range}
		& \textbf{GRAVITY+ ($100\,\mu\mathrm{as}$)}
		& \textbf{KBI ($10\,\mu\mathrm{as}$)}
		& \textbf{KBI ($1\,\mu\mathrm{as}$)}
		& \textbf{EHT} \citep{2022ApJ...930L..17E} \\
		\hline
		
		Kerr
		& \(0 \leq \chi \leq 1\)
		& \(0 \leq \chi \leq 0.58\)
		& \(0 \leq \chi \leq 0.058\)
		& \(0 \leq \chi \leq 0.0058\)
		& --- \\
		
		Kerr--Newman (\(\bar{q}=0.25\))
		& \(0 \leq \chi \leq \sqrt{1-\bar{q}^{\,2}}\)
		& \(0 \leq \chi \leq 0.40\)
		& \(\times\)
		& \(\times\)
		& --- \\
		
		Kerr--Newman (\(\bar{q}=0.95\))
		& \(0 \leq \chi \leq \sqrt{1-\bar{q}^{\,2}}\)
		& \(\times\)
		& \(\times\)
		& \(\times\)
		& \(\times\) \\
		
		Bardeen (\(\bar{g}=0.25\))
		& \(0 \leq \chi \leq \chi_{\rm max}(\bar{g})\) \(\,^{b}\)
		& \(0 \leq \chi \leq 0.58\)
		& \(0 \leq \chi \leq 0.056\)
		& \(0 \leq \chi \leq 0.0029\)
		& --- \\
		
		Bardeen (\(\bar{g}=0.75\))
		& \(0 \leq \chi \leq \chi_{\rm max}(\bar{g})\) \(\,^{b}\)
		& ---
		& \(0 \leq \chi \leq 0.032\)
		& \(\times\) 
		& --- \\
		
		Hayward (\(\bar{l}=0.25\))
		& \(0 \leq \chi \leq \chi_{\rm max}(\bar{l})\) \(\,^{b}\)
		& \(0 \leq \chi \leq 0.59\)
		& \(0 \leq \chi \leq 0.059\)
		& \(0 \leq \chi \leq 0.0059\)
		& --- \\
		
		Hayward (\(\bar{l}=0.75\))
		& \(0 \leq \chi \leq \chi_{\rm max}(\bar{l})\) \(\,^{b}\)
		& ---
		& \(0 \leq \chi \leq 0.059\)
		& \(0 \leq \chi \leq 0.0057\)
		& --- \\
		
		Sen (\(\bar{q}_{\rm m}=0.25\))
		& \(0 \leq \chi \leq 1-\bar{q}_{\rm m}^{\,2}/2\)
		& \(0 \leq \chi \leq 0.40\)
		& \(\times\)
		& \(\times\)
		& --- \\
		
		Sen (\(\bar{q}_{\rm m}=0.95\))
		& \(0 \leq \chi \leq 1-\bar{q}_{\rm m}^{\,2}/2\)
		& \(\times\)
		& \(\times\)
		& \(\times\)
		& \(\times\) \\
		
		Sen (\(\bar{q}_{\rm m}=1.25\))
		& \(0 \leq \chi \leq 1-\bar{q}_{\rm m}^{\,2}/2\)
		& \(\times\)
		& \(\times\)
		& \(\times\)
		& \(\times\) \\
		
		\hline
		\multicolumn{6}{l}{\footnotesize
			\(^{b}\) The maximum spin is obtained from the corresponding degenerate double horizon condition.}
	\end{tabular*}
\end{table*}

\emph{\textbf{Observational forecast.---}}
Current observations of S2 are not sensitive to the spin of Sgr~A* \citep{2024A&A...692A.242G}. S301 provides a promising target for such measurements \citep{2026Natur.657..359A}. S301 has an orbital period of \(8.68\,{\rm yr}\), a semi-major axis of \(83\,{\rm mas}\), and an eccentricity\footnote{S301 admits two possible orbital solutions \cite[cf.][]{2026Natur.657..359A}. We adopt the solution that provides a conservative forecast.} \(e=0.9821\), and is expected to be sensitive to the black-hole spin at an astrometric precision of 100 µas, representative of the expected final performance of GRAVITY+ \citep{2026Natur.657..359A}. It therefore provides an opportunity to probe deviations from the Schwarzschild spacetime.

Further sensitivity can be gained through improved astrometric precision. The proposed kilometre-baseline interferometer (KBI) concept is expected to reach precisions in the \(1\)--\(10\,\mu{\rm as}\) range \citep{eisenhauer2026kbi, 2026A&A...705A.216W}, extending the parameter space accessible with stellar orbits around Sgr~A*. We therefore consider forecasts for both the expected GRAVITY+ precision and the prospective KBI precision.

To describe the projected orbit, we introduce the complex sky coordinate \citep[e.g.][]{2022A&A...667A.172D}
\begin{equation}
	\xi=x+iy,
\end{equation}
which can be written as
\begin{equation}
	\xi
	=
	r e^{i\Omega_\star}
	\left[
	\cos^2\left(\frac{I_\star}{2}\right)e^{iu}
	+
	\sin^2\left(\frac{I_\star}{2}\right)e^{-iu}
	\right],
	\label{eq:skyposition}
\end{equation}
where \(r=a(1-e\cos E)\), \(E\) is the eccentric anomaly, \(u=\nu+\omega_\star\), and \(\nu\) is the true anomaly. The angles \(\Omega_\star\), \(\omega_\star\), and \(I_\star\) denote, respectively, the longitude of the ascending node, argument of pericentre, and inclination, all defined in the observer’s frame\footnote{The starred quantities can be found in \citet{2026Natur.657..359A}.}.
When \(r\) is expressed in \({\rm mas}\), \(\xi\) directly gives the angular position on the sky.

We model the evolution of the orbital orientation using a linear secular approximation. To this end, it is useful to consider the in-plane precession,
\begin{equation}
	\dot{\psi}
	=
	\dot{\omega}
	+
	\dot{\Omega}\cos I
	=
	\dot{\omega}_\star
	+
	\dot{\Omega}_\star \cos I_\star,
    \label{ipprec}
\end{equation}
which is invariant under a fixed rotation of the reference frame \citep[e.g.][]{2020ApJ...888L...8N, 2023ApJ...959...58W}.

For nonrotating spacetimes, the orbital plane remains fixed, such that \(\dot{\Omega}=\dot{\Omega}_\star=\dot{I}_\star=0\). The in-plane apsidal precession (Eq.\,(\ref{ipprec})) therefore reduces to \(\dot{\psi}=\dot{\omega}=\dot{\omega}_\star\), and the argument of pericentre evolves as
\begin{equation}
	\omega_\star(t)
	=
	\omega_{\star,0}
	+\dot{\psi}\,t,
	\label{eq:omega-star}
\end{equation}
where \(\omega_{\star,0}\) is the argument of pericentre at the reference time \(t_0\), which we set to \(t_0=0\) for simplicity.

Rotating spacetimes can also induce secular changes in \(\Omega\), \(\Omega_\star\) and \(I_\star\). To maximise the spin-induced in-plane precession, we take the black-hole spin to be aligned with the orbital angular momentum, \(\cos I=1\). The orbital plane then remains fixed in the observer's frame, such that \(\dot{\Omega}_\star=\dot{I}_\star=0\), and the in-plane precession reduces to
\begin{equation}
	\dot{\psi}
	=
	\dot{\omega}
	+
	\dot{\Omega}
	=
	\dot{\omega}_\star,
	\label{eq:psi-aligned}
\end{equation}
and follows the same evolution as in Eq.~\eqref{eq:omega-star}.

We quantify the projected secular displacement between the Schwarzschild and alternative-spacetime predictions as
\begin{equation}
	\Delta\Theta(t)
	\equiv
	\left|\xi_{\rm S}(t)-\xi_{\rm alt}(t)\right|.
	\label{eq:angular-residual}
\end{equation}
For a fixed orbital plane, Eq.~(\ref{eq:skyposition}) gives, to leading order in \(\delta\psi\),
\begin{align}
	\Delta\Theta
	={}&
	r
	\left|\delta\psi\right|
	\sqrt{
		1-\sin^2 I_\star
		\cos^2 u_{\rm S}
	},
	\label{eq:angular-residual-explicit}
\end{align}
where
\begin{equation}
	\delta\psi
	=
	\left(
	\dot{\psi}_{\rm alt}
	-
	\dot{\psi}_{\rm S}
	\right)t,
	\label{eq:deltapsi}
\end{equation}
and
\begin{equation}
	u_{\rm S}
	=
	\nu
	+
	\omega_{\star,0}
	+
	\dot{\psi}_{\rm S}t.
	\label{eq:u-gr}
\end{equation}

Since the projected displacement varies along the orbit, we characterise its detectability over a given observing baseline \(T_{\rm obs}\). Using the S301 data \citep{2026Natur.657..359A}, we adopt \(T_{\rm obs}=P=8.68\,\mathrm{yr}\), and consider the maximum deviation over the orbit,
\begin{equation}
	\Delta\Theta_{\rm max}
	\simeq
	C_{\rm max}\left|\Delta\psi\right|,
	\label{eq:maximum-residual}
\end{equation}
where \(C_{\rm max}=9.37\times10^4\,\mu\mathrm{as}\,\mathrm{rad}^{-1}\), and
\begin{equation}
	\left|\Delta\psi\right|
	=
	\left|
	\Delta\psi_{\rm alt}
	-
	\Delta\psi_{\rm S}
	\right|,
\end{equation}
the corresponding values are given in Table~\ref{tab:hamiltonian_precession}, while \(\Delta\psi_{\rm S} = 0.0327\,\mathrm{rad}\) (Eq.~\eqref{eq:schwarzschildcheck}).

As an idealised sensitivity criterion~\footnote{This criterion neglects orbital-element covariances and astrophysical/instrumental systematics, and therefore provides an idealised sensitivity estimate. A complete parameter-estimation analysis lies beyond the scope of this work.}, 
we regard a deviation as potentially resolvable when
\(\Delta\Theta_{\rm max}\gtrsim\sigma_{\rm ast}\), where
\(\sigma_{\rm ast}\) denotes the astrometric precision. We use this criterion to define forecast sensitivity limits on the corresponding alternative-spacetime parameters.

Independent constraints on the spacetime of Sgr~A* have been obtained by the Event Horizon Telescope (EHT) through observations of its shadow \citep{2022ApJ...930L..17E}. We consider both spacetimes already constrained by the EHT to determine whether stellar astrometry can provide tighter bounds, and those that remain unconstrained, for which observations of S301 with GRAVITY+ or KBI may provide complementary constraints.

\emph{\textbf{Results and discussion.---}}
In Fig.~\ref{fig:deviations_non_rotating}, we show the maximum astrometric residual accumulated over one orbit of S301 between the Schwarzschild spacetime and the nonrotating spacetimes considered by \citet{2022ApJ...930L..17E}, together with the Simpson--Visser geometry \citep{2019JCAP...02..042S}. At the expected astrometric precision of GRAVITY+, single-orbit S301 monitoring would be sensitive to the Reissner--Nordstr\"om (RN), Einstein--Maxwell--dilaton (EMd-1), Frolov with \(\bar{\ell}=0.4\), and Kazakov \& Solodukhin (KS) spacetimes. At the least precise KBI performance (\(10\mu as\)), these limits improve and extend to the Simpson--Visser spacetime. 
At the best KBI precision (\(1\mu as\)), the allowed parameter ranges for all of these geometries are further reduced, and the Bardeen spacetime would also be detectable. The Hayward spacetime falls below the sensitivity threshold within a single S301 orbit at all astrometric precisions considered. Nevertheless, because secular residuals accumulate over time, multi-orbit monitoring could in principle bring the Hayward deviations within reach of both KBI and GRAVITY+.

Table~\ref{tab:constraints_no_rotating} summarises the forecast limits for GRAVITY+ and KBI and compares them with those obtained from the EHT shadow analysis of \citet{2022ApJ...930L..17E}. Stellar astrometry tightens forecast limits on the RN, EMd-1, and KS spacetimes and provides sensitivity to cases for which the EHT analysis yielded no constraint, including the Bardeen and Frolov geometries. These results illustrate the complementary power of S301 astrometry for testing the spacetime of Sgr~A*.

In Fig.~\ref{fig:deviations_rotating}, we present the maximum astrometric residuals for the rotating spacetimes considered by \citet{2022ApJ...930L..17E}. At the precision of GRAVITY+, single-orbit S301 monitoring would be sensitive to the selected Sen solutions with \(\bar{q}_{\rm m}=1.25\) and \(0.95\), and the Kerr--Newman solution with \(\bar{q}=0.95\), over the full range of spins considered. GRAVITY+ is sensitive to Hayward and Bardeen geometries with \(\bar{\ell}=0.25\) and \(\bar{g}=0.25\), respectively, as well as to the Kerr--Newman solution with \(\bar{q}=0.25\), the Sen solution with \(\bar{q}_{\rm m}=0.25\), and the Kerr spacetime. The improved astrometric precision expected from KBI extends the range of detectable spin-induced deviations, allowing forecast sensitivity to extend to substantially lower black-hole spins. In particular, KBI would probe Hayward geometries with \(\bar{\ell}=0.75\) and Bardeen geometries with \(\bar{g}=0.75\). Table~\ref{tab:constraints_rotating} summarises these limits and compares them with the EHT bounds reported by \citet{2022ApJ...930L..17E}.

\emph{\textbf{Conclusions.---}}
We have developed a Hamiltonian framework for computing the secular apsidal and nodal precessions of bound orbits in stationary, axisymmetric spacetimes admitting a post-Newtonian expansion. Starting from the coordinate-time particle Hamiltonian, a near-identity Lie transformation yields the secular Hamiltonian, from which the accumulated precessions follow directly in Delaunay variables. The framework can be systematically extended to additional stationary and axisymmetric spacetimes, higher post-Newtonian orders, and other secular orbital observables.

We applied this framework to forecast S301’s sensitivity to departures from Schwarzschild spacetime. We evolved the resulting astrometric residual secularly over one orbit and compared its maximum value with \(100\,\mu\mathrm{as}\), representative of the expected GRAVITY+ performance, and with the prospective KBI precisions.
At the expected GRAVITY+ precision, S301 could improve existing constraints on several nonrotating spacetimes and provide sensitivity to geometries that were not constrained by the EHT (cf. Tables~\ref{tab:constraints_no_rotating} and ~\ref{tab:constraints_rotating}), including the Frolov spacetime with \(\bar{\ell}=0.4\). For rotating spacetimes, S301 could probe parameter ranges beyond those accessible to the EHT and provide sensitivity where shadow observations yield no constraints. The improved precisions expected from KBI could tighten these limits and extend the sensitivity to the Simpson--Visser and Bardeen spacetimes. In the most favourable cases, the forecast deformation limits lie very close to the Schwarzschild limit, while for rotating geometries the forecast spin range could be restricted to values close to zero if no deviation from Schwarzschild is detected.

For the rotating spacetimes, we have assumed that the spin axis of Sgr~A* is aligned with the orbital angular momentum of S301, which maximises the spin-induced projected astrometric signal. A generic misalignment may reduce this signal and partially mask deviations from Schwarzschild spacetime. Constraining the full spin vector of Sgr~A* will therefore be important for distinguishing the Kerr solution from alternative rotating geometries such as those considered here and in \citet{2022ApJ...930L..17E}.

Our results establish that S301 is a sensitive probe of the spacetime geometry of Sgr~A*. The secular framework developed here establishes a general mapping from weak-field deformations of stationary, axisymmetric metrics to measurable stellar astrometric observables. Improved KBI astrometry and the discovery of even more compact stars would further strengthen these tests of the Schwarzschild and Kerr paradigms.

\emph{\textbf{Acknowledgements.---}}
This work was financed through national funds by FCT---Funda\c{c}\~ao para a Ci\^encia e a Tecnologia, I.P., Portugal, through projects \href{https://doi.org/10.54499/2024.01252.BD}{2024.01252.BD} and CFisUC UID/04564/2025 (DOI: \href{https://doi.org/10.54499/UID/04564/2025}{10.54499/UID/04564/2025}), and by the Center for Astrophysics and Gravitation (CENTRA/IST/ULisboa) through grants UID/PRR/00099/2025 (DOI: \href{https://doi.org/10.54499/UID/PRR/00099/2025}{10.54499/UID/PRR/00099/2025}) and UID/00099/2025 (DOI: \href{https://doi.org/10.54499/UID/00099/2025}{10.54499/UID/00099/2025}).



\bibliography{prl}

\clearpage
\onecolumngrid

\makeatletter
\let\set@footnotewidth\set@footnotewidth@one
\makeatother


\begin{center}
	{\large\bfseries
		Supplemental Material for:\\[0.3em]
		Probing departures from Schwarzschild geometry with the S301 star and kilometre-baseline interferometry
	}
	
	\vspace{1em}
	
	Rodrigo P. Silva,$^{1,2,3}$
	Alexandre C. M. Correia,$^{1,4}$
	and Paulo J. V. Garcia$^{2,3}$
	
	\vspace{0.5em}
	
	{\small
		$^1$CFisUC, Departamento de F\'isica, Universidade de Coimbra,
		3004-516 Coimbra, Portugal\\
		$^2$Faculdade de Engenharia, Universidade do Porto,
		Rua Dr. Roberto Frias, 4200-465 Porto, Portugal\\
		$^3$CENTRA---Centro de Astrof\'isica e Gravita\c{c}\~ao,
		IST, Universidade de Lisboa, 1049-001 Lisboa, Portugal\\
		$^4$LTE, Observatoire de Paris, Universit\'e PSL,
		Sorbonne Universit\'e, CNRS, 75014 Paris, France
	}
\end{center}

\vspace{1em}

\setcounter{secnumdepth}{1}

\setcounter{section}{0}
\renewcommand{\thesection}{S\Roman{section}}

\counterwithout{equation}{section}
\setcounter{equation}{0}
\renewcommand{\theequation}{S\arabic{equation}}

\section{Coordinate-time particle Hamiltonian}
\label{app:hamiltonian_derivation}

We consider a stationary spacetime written in \(3+1\) form \citep[e.g.][]{2007gr.qc.....3035G, 2010grav.book.....P},
\begin{equation}
	ds^2
	=
	-\alpha^2c^2dt^2
	+
	\gamma_{ij}
	\left(dq^i+c\beta^i dt\right)
	\left(dq^j+c\beta^j dt\right),
	\label{eq:sm_31_metric}
\end{equation}
where \(c\) is the speed of light, \(\alpha\) is the lapse function, \(\beta^i\) is the shift vector, \(\gamma_{ij}\) is the spatial metric, \(t\) is the time coordinate, and \(q^i\) are the spatial coordinates, with \(i,j=1,2,3\).

The action of a test particle of mass \(m\) is \citep[e.g.][]{2014grav.book.....P}
\begin{equation}
	\mathcal S
	=
	-mc^2\int d\tau
	=
	\int \mathcal L\,dt,
\end{equation}
where \(\mathcal L\) is the Lagrangian.

Using \(c^2d\tau^2=-ds^2\), we have
\begin{equation}
	\mathcal L
	=
	-mc^2
	\left[
	\alpha^2
	-
	\frac{\gamma_{ij}}{c^2}
	\left(\dot q^i+c\beta^i\right)
	\left(\dot q^j+c\beta^j\right)
	\right]^{1/2},
	\label{eq:sm_lagrangian}
\end{equation}
where \(\dot q^i\equiv dq^i/dt\). Defining
\begin{equation}
	\Psi
	\equiv
	\frac{d\tau}{dt}
	=
	\left[
	\alpha^2
	-
	\frac{\gamma_{ij}}{c^2}
	\left(\dot q^i+c\beta^i\right)
	\left(\dot q^j+c\beta^j\right)
	\right]^{1/2},
\end{equation}
the canonical momenta are
\begin{equation}
	p_i
	=
	\frac{\partial\mathcal L}{\partial\dot q^i}
	=
	\frac{m}{\Psi}
	\gamma_{ij}
	\left(\dot q^j+c\beta^j\right).
	\label{eq:sm_momenta}
\end{equation}
Contracting Eq.~\eqref{eq:sm_momenta} with the inverse spatial metric and using the definition of \(\Psi\) gives
\begin{equation}
	\Psi^2
	=
	\frac{m^2c^2\alpha^2}
	{m^2c^2+\gamma^{ij}p_ip_j}.
	\label{eq:sm_psi}
\end{equation}
Equation~\eqref{eq:sm_momenta} also gives
\begin{equation}
	\dot q^i
	=
	\frac{\Psi}{m}\gamma^{ij}p_j
	-c\beta^i .
\end{equation}
The Legendre transformation then yields
\begin{align}
	\mathcal H
	&=
	p_i\dot q^i-\mathcal L
	\nonumber\\
	&=
	\frac{\Psi}{m}
	\left(
	m^2c^2+\gamma^{ij}p_ip_j
	\right)
	-c\beta^ip_i
	\nonumber\\
	&=
	\alpha c
	\sqrt{m^2c^2+\gamma^{ij}p_ip_j}
	-
	c\beta^ip_i.
	\label{eq:sm_exact_hamiltonian}
\end{align}
Equation~\eqref{eq:sm_exact_hamiltonian} generates the coordinate-time evolution of a massive test particle in the chosen \(3+1\) foliation. For a stationary spacetime, \(\mathcal H\) has no explicit dependence on \(t\).


\section{Post-Newtonian expansion and secular averaging}
\label{app:general_pn}
We consider asymptotically flat, stationary, axisymmetric spacetimes that admit a weak-field post-Newtonian (PN) expansion and recover a Keplerian Newtonian limit. We expand the \(3+1\) fields as \citep{1997PThPS.128..123A}
\begin{align}
	\alpha
	&=
	1+\alpha_{(2)}
	+\alpha_{(4)}
	+\alpha_{(6)}
	+\mathcal O(\epsilon^{8}),
	\label{eq:sm_field_expansions_1}
	\\
	\beta^i
	&=
	\beta^i_{(4)}
	+\mathcal O(\epsilon^{6}),
	\label{eq:sm_field_expansions_2}
	\\
	\gamma^{ij}
	&=
	h^{ij}
	+\sigma^{ij}_{(2)}
	+\sigma^{ij}_{(4)}
	+\mathcal O(\epsilon^{6}),
	\label{eq:sm_field_expansions_3}
\end{align}
where the subscript denotes the PN scaling of the corresponding \(3+1\) field, such that
\(\alpha_{(n)}=\mathcal O(\epsilon^n)\) and
\(\sigma^{ij}_{(n)}=\mathcal O(\epsilon^n)\), with
\begin{equation}
	\epsilon^2
	\sim
	\frac{v^2}{c^2}
	\sim
	\frac{\mu}{rc^2},
	\qquad
	\mu\equiv\mathcal G M_\bullet,
\end{equation}
where \(\mathcal G\) is the gravitational constant and \(M_\bullet\) is the mass of Sgr~A*.
For the spin sector, we adopt the counting appropriate to a fixed dimensionless black-hole spin,
\begin{equation}
	\chi
	\equiv
	\frac{cJ_\bullet}{\mathcal G M_\bullet^2},
\end{equation}
where \(J_\bullet\) denotes the black-hole spin and for which \(\beta^i_{(4)}=\mathcal O(\epsilon^4)\).
In Eq.~\eqref{eq:sm_field_expansions_3}, \(h^{ij}\) is the inverse Euclidean metric in the chosen coordinate system, and \(\sigma^{ij}_{(n)}\) denotes the deviation of the inverse spatial metric from its Euclidean value at the corresponding PN order. The Keplerian Newtonian limit fixes
\begin{equation}
	\alpha_{(2)}
	=
	-\frac{\mu}{rc^2}.
\end{equation}

Defining
\begin{equation}
	p^2\equiv h^{ij}p_ip_j,
\end{equation}
and subtracting the constant rest-mass contribution, we write the Hamiltonian through 2PN order as
\begin{equation}
	\mathcal H^\prime
	=
	\mathcal H_0
	+\epsilon^2\mathcal H_1
	+\epsilon^3\mathcal H_{1.5}
	+\epsilon^4\mathcal H_2
	+\mathcal O(\epsilon^5),
	\label{eq:sm_pn_hamiltonian}
\end{equation}
where 
\begin{equation}
	\mathcal H_0
	=
	\frac{p^2}{2m}
	-
	\frac{\mu m}{r}
	\label{eq:sm_kepler_hamiltonian}
\end{equation}
is the Kepler Hamiltonian. The PN contributions appearing in Eq.~\eqref{eq:sm_pn_hamiltonian} are
\begin{align}
	\epsilon^2\mathcal H_1
	={}&
	-\frac{p^4}{8m^3c^2}
	+\frac{\alpha_{(2)}p^2}{2m}
	+\frac{\sigma^{ij}_{(2)}p_ip_j}{2m}
	+mc^2\alpha_{(4)},
	\nonumber\\
	\epsilon^3\mathcal H_{1.5}
	={}&
	-c\,\beta^i_{(4)}p_i,
	\nonumber\\
	\epsilon^4\mathcal H_2
	={}&
	\frac{p^6}{16m^5c^4}
	-\frac{p^2\sigma^{ij}_{(2)}p_ip_j}{4m^3c^2}
	-\frac{\alpha_{(2)}p^4}{8m^3c^2}
	+\frac{\alpha_{(2)}
		\sigma^{ij}_{(2)}p_ip_j}{2m}
	\nonumber\\
	&+
	\frac{\alpha_{(4)}p^2}{2m}
	+\frac{\sigma^{ij}_{(4)}p_ip_j}{2m}
	+mc^2\alpha_{(6)}.
	\label{eq:sm_pn_components}
\end{align}
The coefficients \(\alpha_{(n)}\), \(\beta^i_{(n)}\), and
\(\sigma^{ij}_{(n)}\) therefore determine the perturbing Hamiltonian at each PN order. If an additional non-Keplerian contribution enters formally at Newtonian order but is controlled by an independent small parameter, we retain the Kepler Hamiltonian \(\mathcal H_0\) as the integrable part and treat that contribution perturbatively. This case is illustrated below for the Simpson--Visser geometry.

\section{Schwarzschild spacetime}
\label{app:schwarzschild}

\subsection{First post-Newtonian order}
The Schwarzschild metric is
\citep{1916SPAW.......189S}
\begin{equation}
	ds^2
	=
	-\left(1-\frac{r_{\rm S}}{r}\right)c^2dt^2
	+
	\left(1-\frac{r_{\rm S}}{r}\right)^{-1}dr^2
	+
	r^2d\Omega^2,
	\label{eq:sm_schwarzschild_metric}
\end{equation}
where
\begin{equation}
	r_{\rm S}=\frac{2\mu}{c^2}.
\end{equation}
Comparing with Eqs.~\eqref{eq:sm_field_expansions_1}--\eqref{eq:sm_field_expansions_3}, the weak-field coefficients relevant at 1PN order are
\begin{equation}
	\alpha_{(2)}
	=
	-\frac{\mu}{rc^2},
	\qquad
	\alpha_{(4)}
	=
	-\frac{\mu^2}{2r^2c^4},
	\qquad
	\sigma^{rr}_{(2)}
	=
	-\frac{2\mu}{rc^2},
	\label{eq:sm_schwarzschild_coefficients}
\end{equation}
while all components of the shift vanish. The 1PN Hamiltonian contribution therefore becomes
\begin{equation}
	\epsilon^2\mathcal H_1
	=
	\frac{1}{c^2}
	\left[
	-\frac{p^4}{8m^3}
	-\frac{\mu p^2}{2mr}
	-\frac{\mu p_r^2}{mr}
	-\frac{\mu^2m}{2r^2}
	\right],
	\label{eq:sm_schwarzschild_h1}
\end{equation}
where
\begin{equation}
	p^2
	=
	p_r^2
	+
	\frac{p_\theta^2}{r^2}
	+
	\frac{p_\phi^2}{r^2\sin^2\theta}.
\end{equation}
By spherical symmetry, the orbital plane may be chosen as the equatorial plane, \(\theta=\pi/2\), for which \(p_\theta=0\) and
\begin{equation}
	p^2
	=
	p_r^2+\frac{G^2}{r^2},
\end{equation}
where \(G\) is the magnitude of the Newtonian orbital angular momentum.

Evaluating the secular average of \(\mathcal H_1\) allows us to write
\begin{equation}
	\langle\epsilon^2\mathcal H_1\rangle
	=
	\frac{1}{c^2}
	\left[
	-\frac{3\mu^2m}{a^2\sqrt{1-e^2}}
	+
	\frac{15\mu^2m}{8a^2}
	\right].
	\label{eq:sm_schwarzschild_h1_average}
\end{equation}
Equivalently, using the Delaunay actions,
\begin{equation}
	a=\frac{L^2}{m^2\mu},
	\qquad
	\sqrt{1-e^2}=\frac{G}{L},
\end{equation}
the averaged 1PN Hamiltonian contribution can be written as
\begin{equation}
	\langle\epsilon^2\mathcal H_1\rangle
	=
	\frac{m^5\mu^4}{c^2}
	\left(
	\frac{15}{8L^4}
	-
	\frac{3}{L^3G}
	\right).
	\label{eq:sm_schwarzschild_h1_delaunay}
\end{equation}
The Kepler Hamiltonian in the same variables is
\begin{equation}
	\mathcal H_0
	=
	-\frac{m^3\mu^2}{2L^2}.
\end{equation}
Expanding Eq.~\eqref{eq:precessions} consistently through 1PN order therefore gives
\begin{equation}
	\Delta\omega_{\rm S}
	=
	\int_{0}^{2\pi/n}\dot{\omega}_{\rm S}\,dt
	=
	\frac{2\pi}{n} 
	\frac{\partial\langle\epsilon^2\mathcal H_1\rangle}
	{\partial G}
	=
	\frac{6\pi\mu}{ac^2(1-e^2)}
	\label{eq:sm_schwarzschild_1pn}
\end{equation}
where, as in Eq.~\eqref{eq:precessions},
\begin{equation}
	n
	\equiv
	\frac{\partial\langle\mathcal H^\prime\rangle}{\partial L},
	\qquad
	\langle\mathcal H^\prime\rangle = H_0 + \langle\epsilon^2\mathcal H_1\rangle.
\end{equation}
The nodal precession is
\begin{equation}
	\Delta\Omega_{\rm S}=0.
\end{equation}
that vanishes because spherical symmetry makes the secular Hamiltonian independent of \(H\).


\section{Second post-Newtonian order}
At 2PN order, the additional Schwarzschild Hamiltonian contribution is
\begin{equation}
	\epsilon^4\mathcal H_2
	=
	\frac{1}{c^4}
	\left[
	\frac{p^6}{16m^5}
	+
	\frac{\mu p^4}{8m^3r}
	+
	\frac{\mu p^2p_r^2}{2m^3r}
	+
	\frac{\mu^2p_r^2}{mr^2}
	-
	\frac{\mu^2p^2}{4mr^2}
	-
	\frac{\mu^3m}{2r^3}
	\right].
	\label{eq:sm_schwarzschild_h2}
\end{equation}
The secular contribution of the direct 2PN term is
\begin{equation}
	\langle\epsilon^4\mathcal H_2\rangle
	=
	\frac{\mu^3m}{a^3c^4}
	\left[
	\frac{15}{16}
	-
	\frac{5}{2\sqrt{1-e^2}}
	+
	\frac{1}{(1-e^2)^{3/2}}
	\right].
	\label{eq:sm_schwarzschild_direct_2pn}
\end{equation}
There is, in addition, an indirect 2PN contribution generated by the canonical transformation.
 
For Schwarzschild, spherical symmetry implies that the 1PN perturbation depends on the orbit only through the fast angle \(\ell\) and the actions \(L\) and \(G\). The Poisson bracket entering the indirect 2PN term therefore reduces to the \((\ell,L)\) canonical pair,
\begin{equation}
	\{\widetilde{\mathcal H}_1,S_1\}
	=
	\frac{\partial\widetilde{\mathcal H}_1}{\partial\ell}
	\frac{\partial S_1}{\partial L}
	-
	\frac{\partial\widetilde{\mathcal H}_1}{\partial L}
	\frac{\partial S_1}{\partial\ell}.
	\label{eq:sm_schwarzschild_indirect_bracket}
\end{equation}
Integrating the first term by parts over the periodic angle \(\ell\) and using \citep[cf.][]{1983rsm..book.....L}
\begin{equation}
	\frac{\partial S_1}{\partial\ell}
	=
	\frac{\widetilde{\mathcal H}_1}{n_0},
	\qquad
	n_0
	\equiv
	\frac{\partial\mathcal H_0}{\partial L}
	=
	\frac{m^3\mu^2}{L^3}.
\end{equation}
we obtain
\begin{align}
	\left\langle\epsilon^4
	\{\widetilde{\mathcal H}_1,S_1\}
	\right\rangle
	={}&
	-\left\langle\epsilon^4
	\widetilde{\mathcal H}_1
	\frac{\partial}{\partial L}
	\left(
	\frac{\partial S_1}{\partial\ell}
	\right)
	\right\rangle
	-
	\left\langle\epsilon^4
	\frac{\partial\widetilde{\mathcal H}_1}{\partial L}
	\frac{\partial S_1}{\partial\ell}
	\right\rangle
	\nonumber\\
	={}&
	-\frac{2}{n_0}
	\left\langle\epsilon^4
	\widetilde{\mathcal H}_1
	\frac{\partial\widetilde{\mathcal H}_1}{\partial L}
	\right\rangle
	+
	\frac{1}{n_0^2}
	\frac{\partial n_0}{\partial L}
	\left\langle\epsilon^4
	\widetilde{\mathcal H}_1^2
	\right\rangle
	\nonumber\\
	={}&
	-\frac{\partial}{\partial L}
	\left[
	\frac{
		\left\langle\epsilon^4\widetilde{\mathcal H}_1^2\right\rangle
	}{n_0}
	\right].
	\label{eq:sm_schwarzschild_bracket_intermediate}
\end{align}

To evaluate the remaining average, we introduce
\begin{equation}
	s
	\equiv
	\sqrt{1-e^2}
	=
	\frac{G}{L},
	\qquad
	D
	\equiv
	1-e\cos E,
	\label{eq:sm_schwarzschild_su}
\end{equation}
so that \(r=aD\) and
\begin{equation}
	p^2
	=
	\frac{m^2\mu}{a}
	\left(
	\frac{2}{D}-1
	\right),
	\qquad
	p_r
	=
	m\sqrt{\frac{\mu}{a}}\,
	\frac{e\sin E}{D}.
	\label{eq:sm_schwarzschild_kepler_2pn}
\end{equation}
Using
\begin{equation}
	e^2\sin^2E
	=
	2D-D^2-s^2,
\end{equation}
the 1PN Hamiltonian in Eq.~\eqref{eq:sm_schwarzschild_h1} can be written as
\begin{equation}
	\epsilon^2\mathcal H_1
	=
	\frac{\mu^2m}{a^2c^2}f(E),
	\qquad
	f(E)
	=
	-\frac{1}{8}
	+\frac{2}{D}
	-\frac{4}{D^2}
	+\frac{s^2}{D^3}.
	\label{eq:sm_schwarzschild_f}
\end{equation}
Since \(d\ell=D\,dE\), the secular average gives
\begin{equation}
	\langle f\rangle
	=
	\frac{1}{2\pi}
	\int_0^{2\pi}
	f(E)D\,dE
	=
	\frac{15}{8}
	-
	\frac{3}{s},
	\label{eq:sm_schwarzschild_f_average}
\end{equation}
in agreement with Eq.~\eqref{eq:sm_schwarzschild_h1_average}. Hence
\begin{equation}
	\epsilon^2\widetilde{\mathcal H}_1
	=
	\frac{\mu^2m}{a^2c^2}
	\left[
	f(E)-\langle f\rangle
	\right].
	\label{eq:sm_schwarzschild_h1_tilde}
\end{equation}
A direct secular average then yields
\begin{align}
	\left\langle\epsilon^4
	\widetilde{\mathcal H}_1^2
	\right\rangle
	&=
	\left(
	\frac{\mu^2m}{a^2c^2}
	\right)^2
	\left[
	\langle f^2\rangle-\langle f\rangle^2
	\right]
	=
	\left(
	\frac{\mu^2m}{a^2c^2}
	\right)^2
	V(s),
	\label{eq:sm_schwarzschild_h1_variance}
\end{align}
where
\begin{equation}
	V(s)
	=
	-4
	+\frac{115}{8s}
	-\frac{9}{s^2}
	-\frac{39}{4s^3}
	+\frac{67}{8s^5}.
	\label{eq:sm_schwarzschild_variance_function}
\end{equation}

Using
\begin{equation}
	a=\frac{L^2}{m^2\mu},
	\qquad
	s=\frac{G}{L},
	\qquad
	n_0=\frac{m^3\mu^2}{L^3},
\end{equation}
we obtain
\begin{equation}
	\frac{
		\langle\epsilon^4\widetilde{\mathcal H}_1^2\rangle
	}{n_0}
	=
	\frac{m^7\mu^6}{L^5c^4}V(s).
\end{equation}
It follows that
\begin{align}
	\frac{1}{2}
	\left\langle\epsilon^4
	\{\widetilde{\mathcal H}_1,S_1\}
	\right\rangle
	&=
	-\frac{1}{2}
	\frac{\partial}{\partial L}
	\left[
	\frac{m^7\mu^6}{L^5c^4}
	V(s)
	\right]
	=
	\frac{m^7\mu^6}{2L^6c^4}
	\left[
	5V(s)+sV'(s)
	\right].
	\label{eq:sm_schwarzschild_indirect_derivative}
\end{align}
From Eq.~\eqref{eq:sm_schwarzschild_variance_function},
\begin{equation}
	\frac{1}{2}
	\left[
	5V(s)+sV'(s)
	\right]
	=
	-10
	+\frac{115}{4s}
	-\frac{27}{2s^2}
	-\frac{39}{4s^3}.
\end{equation}
Finally, using
\begin{equation}
	\frac{m^7\mu^6}{L^6c^4}
	=
	\frac{\mu^3m}{a^3c^4},
\end{equation}
and restoring \(s=\sqrt{1-e^2}\), the indirect 2PN contribution is
\begin{equation}
	\frac{1}{2}
	\left\langle\epsilon^4
	\{\widetilde{\mathcal H}_1,S_1\}
	\right\rangle
	=
	\frac{\mu^3m}{a^3c^4}
	\left[
	-10
	+
	\frac{115}{4\sqrt{1-e^2}}
	-
	\frac{27}{2(1-e^2)}
	-
	\frac{39}{4(1-e^2)^{3/2}}
	\right].
	\label{eq:sm_schwarzschild_indirect_2pn}
\end{equation}

Combining the direct and indirect contributions, we define the complete secular coefficient at 2PN order as
\begin{align}
	\mathcal K_2
	&\equiv
	\langle\epsilon^4\mathcal H_2\rangle
	+
	\frac{1}{2}
	\left\langle\epsilon^4
	\{\widetilde{\mathcal H}_1,S_1\}
	\right\rangle,
\end{align}
such that
\begin{equation}
	\mathcal K_2
	=
	\frac{\mu^3m}{a^3c^4}
	\left[
	-\frac{145}{16}
	+
	\frac{105}{4\sqrt{1-e^2}}
	-
	\frac{27}{2(1-e^2)}
	-
	\frac{35}{4(1-e^2)^{3/2}}
	\right].
	\label{eq:sm_schwarzschild_full_2pn}
\end{equation}

In terms of the Delaunay actions,
\begin{align}
	\langle\epsilon^2\mathcal H_1\rangle
	={}&
	\frac{m^5\mu^4}{c^2}
	\left(
	\frac{15}{8L^4}
	-
	\frac{3}{L^3G}
	\right),
	\\
	\mathcal K_2
	={}&
	\frac{m^7\mu^6}{c^4}
	\left(
	-\frac{145}{16L^6}
	+\frac{105}{4L^5G}
	-\frac{27}{2L^4G^2}
	-\frac{35}{4L^3G^3}
	\right).
	\label{eq:sm_schwarzschild_2pn_delaunay}
\end{align}
The secular Hamiltonian through 2PN order is therefore
\begin{equation}
	\langle\mathcal H^\prime\rangle
	=
	\mathcal H_0
	+
	\langle\epsilon^2\mathcal H_1\rangle
	+
	\mathcal K_2
	\label{eq:sm_schwarzschild_secular_2pn}
\end{equation}
Then, the apsidal precession through 2PN order is
\begin{equation}
	\Delta\omega_{\rm S}
	=
	\frac{2\pi}{n}
	\frac{\partial\langle\mathcal H^\prime\rangle}{\partial G}
	=
	\frac{6\pi\mu}
	{ac^2(1-e^2)}
	+
	\frac{15\pi\mu^2(6+e^2)}
	{2a^2c^4(1-e^2)^2}
	\label{eq:sm_schwarzschild_2pn}
\end{equation}
where \(a\) and \(e\) denote the semi-major axis and eccentricity of the orbit, respectively. We recover the 2PN result reported in the literature \citep[e.g.][]{1988NCimB.101..127D}. 

Finally,
\begin{equation}
	\Delta\Omega_{\rm S}=0,
\end{equation}
since spherical symmetry fixes the orbital plane and therefore forbids nodal precession.

\section{Kerr spacetime}
\label{app:kerr}

The Kerr metric in Boyer--Lindquist coordinates is
\citep{1963PhRvL..11..237K,1967JMP.....8..265B}
\begin{align}
	ds^2
	={}&
	-\left(1-\frac{r_{\rm S}r}{\Sigma}\right)c^2dt^2
	-
	\frac{2a_\bullet r_{\rm S}r\sin^2\theta}{\Sigma}
	c\,dt\,d\phi
	\nonumber\\
	&+
	\frac{\Sigma}{\Delta}\,dr^2
	+
	\Sigma\,d\theta^2
	+
	\frac{A\sin^2\theta}{\Sigma}\,d\phi^2,
	\label{eq:sm_kerr_metric}
\end{align}
where
\begin{align}
	\Sigma
	&=
	r^2+a_\bullet^2\cos^2\theta,
	\nonumber\\
	\Delta
	&=
	r^2-r_{\rm S}r+a_\bullet^2,
	\nonumber\\
	A
	&=
	(r^2+a_\bullet^2)^2
	-
	a_\bullet^2\Delta\sin^2\theta,
\end{align}
and
\begin{equation}
	a_\bullet
	=
	\frac{J_\bullet}{M_\bullet c},
	\qquad
	J_\bullet
	=
	\frac{\chi\mu^2}{c\mathcal G}.
	\label{eq:sm_kerr_spin}
\end{equation}

To isolate the leading spin effect, we retain the Schwarzschild 1PN contribution together with the leading 1.5PN spin--orbit term and work to linear order in \(\chi\). The corresponding component of the shift has the weak-field expansion
\begin{equation}
	\beta^\phi_{(4)}
	=
	-\frac{2\chi\mu^2}{c^4r^3}
	\label{eq:sm_kerr_shift}
\end{equation}
Using the general relation
\begin{equation}
	\epsilon^3\mathcal H_{1.5}
	=
	-c\,\beta^i_{(4)}p_i,
\end{equation}
the leading spin--orbit Hamiltonian contribution is therefore
\begin{equation}
	\epsilon^3\mathcal H_{1.5}
	=
	\frac{2\chi\mu^2p_\phi}{c^3r^3}.
	\label{eq:sm_kerr_spin_hamiltonian}
\end{equation}

Because the Kerr symmetry axis defines the polar axis, the conserved azimuthal momentum is
\begin{equation}
	p_\phi
	=
	H
	=
	G\cos I,
\end{equation}
where \(I\) is the inclination of the orbital angular momentum with respect to the black-hole spin axis. Computing the secular average gives us,
\begin{equation}
	\left\langle\frac{1}{r^3}\right\rangle
	=
	\frac{1}{a^3(1-e^2)^{3/2}},
\end{equation}
and hence
\begin{equation}
	\langle\epsilon^3\mathcal H_{1.5}\rangle
	=
	\frac{2\chi m\mu^{5/2}\cos I}
	{c^3a^{5/2}(1-e^2)}.
	\label{eq:sm_kerr_spin_average}
\end{equation}
For taking derivatives with respect to the Delaunay actions, it is useful to rewrite this result using
\begin{equation}
	a=\frac{L^2}{m^2\mu},
	\qquad
	1-e^2=\frac{G^2}{L^2},
	\qquad
	\cos I=\frac{H}{G},
\end{equation}
which gives
\begin{equation}
	\langle\epsilon^3\mathcal H_{1.5}\rangle
	=
	\frac{2\chi m^6\mu^5 H}{c^3L^3G^3}.
	\label{eq:sm_kerr_spin_average_delaunay}
\end{equation}

Retaining the Schwarzschild contribution through 1PN order and the spin contribution through 1.5PN order, the secular Hamiltonian is \footnotemark[2]
\footnotetext[2]{%
	\parbox[t]{0.95\textwidth}{%
		We neglect higher-order terms, since our aim is
		to recover the dominant contributions to the apsidal and nodal
		precessions reported in the literature.
	}%
}
\begin{equation}
	\langle\mathcal H^\prime\rangle
	=
	\mathcal H_0
	+
	\langle\epsilon^2\mathcal H_1\rangle
	+
	\langle\epsilon^3\mathcal H_{1.5}\rangle
	\label{eq:sm_kerr_secular_hamiltonian}
\end{equation}

The precessions accumulated over one period are therefore
\begin{align}
	\Delta\omega
	={}&
	\frac{2\pi}{n}
	\frac{\partial\langle\mathcal H^\prime\rangle}{\partial G}
	=
	\frac{6\pi\mu}{ac^2(1-e^2)}
	-
	\frac{12\pi\chi\mu^{3/2}\cos I}
	{c^3[a(1-e^2)]^{3/2}},
	\nonumber\\
	\Delta\Omega
	={}&
	\frac{2\pi}{n}
	\frac{\partial\langle\mathcal H^\prime\rangle}{\partial H}
	=
	\frac{4\pi\chi\mu^{3/2}}
	{c^3[a(1-e^2)]^{3/2}}.
	\label{eq:sm_kerr_precessions}
\end{align}
The spin-dependent terms satisfy
\begin{equation}
	\Delta\omega_{\rm LT}
	=
	-3\Delta\Omega_{\rm LT}\,\cos I,
\end{equation}
and reproduce the standard leading Lense--Thirring precessions \citep[e.g.][]{2013degn.book.....M}. In the non-spinning limit \(\chi\rightarrow0\), the nodal precession vanishes, and the Schwarzschild 1PN apsidal precession is recovered.

\section{Simpson--Visser spacetime}
\label{app:simpson_visser}

The Simpson--Visser geometry is
\citep{2019JCAP...02..042S}
\begin{align}
	ds^2
	={}&
	-\left(
	1-\frac{r_{\rm S}}{\sqrt{r^2+\lambda^2}}
	\right)c^2dt^2
	+
	\left(
	1-\frac{r_{\rm S}}{\sqrt{r^2+\lambda^2}}
	\right)^{-1}dr^2
	\nonumber\\
	&+
	(r^2+\lambda^2)d\Omega^2,
	\label{eq:sm_simpson_visser_metric}
\end{align}
where \(\lambda\) is the bounce parameter. The cases \(\lambda=0\), \(0<\lambda<r_{\rm S}\), \(\lambda=r_{\rm S}\), and \(\lambda>r_{\rm S}\) correspond, respectively, to Schwarzschild, a regular black-bounce geometry, a one-way wormhole, and a traversable wormhole.

For \(r\gg r_{\rm S}\) and \(r\gg\lambda\), the leading deviation from Schwarzschild is quadratic in \(\lambda\). In particular,
\begin{equation}
	\frac{1}{\sqrt{r^2+\lambda^2}}
	=
	\frac{1}{r}
	-\frac{\lambda^2}{2r^3}
	+\mathcal O\left(\frac{\lambda^4}{r^5}\right),
\end{equation}
while
\begin{equation}
	\frac{1}{r^2+\lambda^2}
	=
	\frac{1}{r^2}
	-\frac{\lambda^2}{r^4}
	+\mathcal O\left(\frac{\lambda^4}{r^6}\right).
\end{equation}
We organise the Hamiltonian as a two-parameter perturbative expansion,
\begin{equation}
	\mathcal H
	=
	\mathcal H_0
	+
	\mathcal H_\lambda
	+
	\epsilon^2\mathcal H_1
	+\cdots,
	\label{eq:sm_simpson_visser_ordering}
\end{equation}
where
\begin{equation}
	\mathcal H_\lambda
	=
	\lambda^2
	\left[
	\frac{\mu m}{2r^3}
	-
	\frac{G^2}{2mr^4}
	\right].
	\label{eq:sm_simpson_visser_perturbation}
\end{equation}
Although \(\mathcal H_\lambda\) enters formally at Newtonian order in the \(1/c\) expansion, it is controlled by the independent small parameter \(\lambda^2/a^2\). We therefore retain the Kepler Hamiltonian \(\mathcal H_0\) as the integrable Hamiltonian and treat \(\mathcal H_\lambda\) perturbatively. We retain terms through first order in \(\lambda^2/a^2\) and through 1PN order, while mixed terms of order \(\epsilon^2\lambda^2/a^2\), terms of order \(\lambda^4/a^4\), and 2PN terms are consistently neglected.

The required secular averages are
\begin{equation}
	\left\langle\frac{1}{r^3}\right\rangle
	=
	\frac{1}{a^3(1-e^2)^{3/2}},
\end{equation}
and
\begin{equation}
	\left\langle\frac{1}{r^4}\right\rangle
	=
	\frac{2+e^2}
	{2a^4(1-e^2)^{5/2}}.
\end{equation}
Using
\begin{equation}
	G^2
	=
	m^2\mu a(1-e^2),
\end{equation}
the secular deformation Hamiltonian becomes
\begin{equation}
	\langle\mathcal H_\lambda\rangle
	=
	-\frac{\lambda^2\mu m}{4a^3}
	\frac{e^2}{(1-e^2)^{3/2}}.
	\label{eq:sm_simpson_visser_average}
\end{equation}
In terms of the Delaunay actions, this can equivalently be written as
\begin{equation}
	\langle\mathcal H_\lambda\rangle
	=
	-\frac{\lambda^2m^7\mu^4}{4}
	\left(
	\frac{1}{L^3G^3}
	-
	\frac{1}{L^5G}
	\right).
	\label{eq:sm_simpson_visser_average_delaunay}
\end{equation}

At the order retained here, the secular Hamiltonian is therefore
\begin{equation}
	\langle\mathcal H^\prime\rangle
	=
	\mathcal H_0
	+
	\langle\mathcal H_\lambda\rangle
	+
	\langle\epsilon^2\mathcal H_1\rangle.
	\label{eq:sm_simpson_visser_secular}
\end{equation}
Since the Kepler Hamiltonian is independent of \(G\), the apsidal precession at zeroth-order is
\begin{equation}
	\Delta\omega_\lambda
	=
	\frac{2\pi}{n}
	\frac{\partial\langle\mathcal H_\lambda\rangle}{\partial G}
	=
	\frac{\pi(2+e^2)\lambda^2}
	{2a^2(1-e^2)^2}.
	\label{eq:sm_simpson_visser_lambda_precession}
\end{equation}

Combining the deformation-induced contribution with the Schwarzschild 1PN precession gives
\begin{equation}
	\Delta\omega
	=
	\frac{6\pi\mu}{c^2a(1-e^2)}
	+
	\frac{\pi(2+e^2)\lambda^2}
	{2a^2(1-e^2)^2}.
	\label{eq:sm_simpson_visser_precession}
\end{equation}
This agrees with Ref.~\cite{Zhou2020PrecessingAP}. Since the Simpson--Visser geometry is spherically symmetric, the secular Hamiltonian is independent of \(H\), and consequently
\begin{equation}
	\Delta\Omega=0.
\end{equation}

\section{Catalogue of spacetime metrics considered}

Here, we catalogue the spacetime metrics considered in this work. The nonrotating spacetimes can be written in the general form \citep[e.g.][]{2001PhRvD..64f4013B}
\begin{equation}
	ds^2
	=
	-f(r)c^2dt^2
	+
	\frac{dr^2}{f(r)}
	+
	R^2(r)\,d\Omega^2,
\end{equation}
where \(d\Omega^2=d\theta^2+\sin^2\theta\,d\phi^2\), \(f(r)\) is the metric function characterising the spacetime, and \(R(r)\) is the areal-radius function. In Table~\ref{tab:nonrotating}, we summarise the normalised parameters for each theory and specify the corresponding forms of \(f(r)\) and \(R(r)\).

\begin{table*}[h]
	\centering
	\caption{Summary of the nonrotating spacetimes considered in this work, their normalised parameters, metric functions, and squared areal-radius functions.}
	\label{tab:nonrotating}
	\footnotesize
	\setlength{\tabcolsep}{5pt}
	
	\begin{tabular}{
			|p{3.0cm}|
			p{3.2cm}|
			p{5.3cm}|
			p{5.4cm}|
		}
		\hline
		
		\rule{0pt}{4ex}\textbf{Spacetime} &
		\textbf{Normalised parameter} &
		\textbf{$f(r)$} &
		\textbf{$R^2(r)$}
		\\[1.5ex]
		\hline
		
		\rule{0pt}{5ex}Reissner--Nordstr\"om \citep{1916AnP...355..106R,1918KNAB...20.1238N}
		&
		$\displaystyle
		\bar q=
		\frac{Q}{M_\bullet\sqrt{4\pi\epsilon_0\mathcal G}}
		$
		&
		$\displaystyle
		1-\frac{2\mu}{c^2r}
		+\frac{\mathcal G Q^2}{4\pi\epsilon_0c^4r^2}
		$
		&
		$\displaystyle
		r^2
		$
		\\[2ex]
		\hline
		
		\rule{0pt}{5ex}Bardeen \citep{1968qtr..conf...87B}
		&
		$\displaystyle
		\bar g=
		\frac{c^2g}{\mu}
		$
		&
		$\displaystyle
		1-\frac{2\mu r^2}
		{c^2(r^2+g^2)^{3/2}}
		$
		&
		$\displaystyle
		r^2
		$
		\\[2ex]
		\hline
		
		\rule{0pt}{5ex}Hayward \citep{2006PhRvL..96c1103H}
		&
		$\displaystyle
		\bar l=
		\frac{c^2l_H}{\mu}
		$
		&
		$\displaystyle
		1-\frac{2(\mu/c^2)r^2}
		{r^3+2(\mu/c^2)l_H^2}
		$
		&
		$\displaystyle
		r^2
		$
		\\[2ex]
		\hline
		
		\rule{0pt}{5ex}Simpson--Visser \citep{2019JCAP...02..042S}
		&
		$\displaystyle
		\bar\lambda=
		\frac{c^2\lambda}{\mu}
		$
		&
		$\displaystyle
		1-\frac{2\mu}
		{c^2\sqrt{r^2+\lambda^2}}
		$
		&
		$\displaystyle
		r^2+\lambda^2
		$
		\\[2ex]
		\hline
		
		\rule{0pt}{8ex}Frolov \((\bar l=0.4)\) \citep{2016PhRvD..94j4056F}
		&
		$\displaystyle
		\begin{aligned}
			\bar q
			&=
			\frac{Q}{M_\bullet\sqrt{4\pi\epsilon_0\mathcal G}},
			\\
			\bar l
			&=
			\frac{c^2l}{\mu}
			=
			0.4
		\end{aligned}
		$
		&
		$\displaystyle
		1-
		\frac{
			\left[
			2(\mu/c^2)r
			-\dfrac{\mathcal G Q^2}{4\pi\epsilon_0c^4}
			\right]r^2
		}{
			r^4+
			\left[
			2(\mu/c^2)r
			+\dfrac{\mathcal G Q^2}{4\pi\epsilon_0c^4}
			\right]l^2
		}
		$
		&
		$\displaystyle
		r^2
		$
		\\[5ex]
		\hline
		
		\rule{0pt}{5ex}KS \citep{1994NuPhB.429..153K}
		&
		$\displaystyle
		\bar\alpha=
		\frac{c^2\alpha}{\mu}
		$
		&
		$\displaystyle
		\sqrt{1-\frac{\alpha^2}{r^2}}
		-\frac{2\mu}{c^2r}
		$
		&
		$\displaystyle
		r^2
		$
		\\[2ex]
		\hline
		
		\rule{0pt}{5ex}EMd-1 \citep{Gibbons:1987ps,Garfinkle:1990qj,1992PhRvD..46.5278K,1992PhRvL..69.1006S}
		&
		$\displaystyle
		\bar q=
		\frac{Q}{M_\bullet\sqrt{4\pi\epsilon_0\mathcal G}}
		$
		&
		$\displaystyle
		1-\frac{2\mu}{c^2r}
		$
		&
		$\displaystyle
		r\left(
		r-\frac{Q^2}{4\pi\epsilon_0M_\bullet c^2}
		\right)
		$
		\\[2ex]
		\hline
		
	\end{tabular}
\end{table*}

The rotating spacetimes considered in this work can be written in the common form \citep[e.g.][]{2014EPJC...74.2865A}
\begin{equation}
	\begin{aligned}
		ds^2
		={}&
		-\left(
		1-\frac{h(r)}{\Sigma(r,\theta)}
		\right)c^2dt^2
		-\frac{2a_\bullet h(r)\sin^2\theta}
		{\Sigma(r,\theta)}\,
		c\,dt\,d\phi
		+\frac{\Sigma(r,\theta)}{\Delta(r)}\,dr^2
		\\
		&+
		\Sigma(r,\theta)\,d\theta^2
		+
		\left[
		R^2(r)+a_\bullet^2
		+\frac{a_\bullet^2h(r)\sin^2\theta}
		{\Sigma(r,\theta)}
		\right]
		\sin^2\theta\,d\phi^2,
	\end{aligned}
\end{equation}
where
\begin{equation}
	\Sigma(r,\theta)
	=
	R^2(r)+a_\bullet^2\cos^2\theta,
	\qquad
	h(r)
	=
	R^2(r)\left[1-f(r)\right],
\end{equation}
and
\begin{equation}
	\Delta(r)
	=
	R^2(r)f(r)+a_\bullet^2.
\end{equation}
In Table~\ref{tab:rotating}, we summarise the rotating spacetimes considered in this work and the functions that define their corresponding metrics.

\begin{table*}[h]
	\centering
	\caption{Summary of the rotating spacetimes considered in this work and the radial functions entering the stationary axisymmetric metric.}
	\label{tab:rotating}
	\footnotesize
	\setlength{\tabcolsep}{4pt}
	
	\begin{tabular}{
			|p{3.0cm}|
			p{3.2cm}|
			p{5.3cm}|
			p{5.4cm}|
		}
		\hline
		
		\rule{0pt}{4ex}
		\textbf{Spacetime}
		&
		\textbf{$R^2(r)$}
		&
		\textbf{$h(r)$}
		&
		\textbf{$\Delta(r)$}
		\\[1.5ex]
		\hline
		
		\rule{0pt}{5ex}
		Kerr \citep{1963PhRvL..11..237K,1967JMP.....8..265B}
		&
		$\displaystyle
		r^2
		$
		&
		$\displaystyle
		\frac{2\mu r}{c^2}
		$
		&
		$\displaystyle
		r^2+a_\bullet^2-\frac{2\mu r}{c^2}
		$
		\\[2ex]
		\hline
		
		\rule{0pt}{6ex}
		Kerr--Newman \citep{1965JMP.....6..918N}
		&
		$\displaystyle
		r^2
		$
		&
		$\displaystyle
		\frac{2\mu r}{c^2}
		-\frac{\mathcal GQ^2}{4\pi\epsilon_0c^4}
		$
		&
		$\displaystyle
		r^2+a_\bullet^2
		-\frac{2\mu r}{c^2}
		+\frac{\mathcal GQ^2}{4\pi\epsilon_0c^4}
		$
		\\[2ex]
		\hline
		
		\rule{0pt}{6ex}
		Bardeen \citep{2016PhRvD..93j4004A}
		&
		$\displaystyle
		r^2
		$
		&
		$\displaystyle
		\frac{2\mu}{c^2}
		\frac{r^4}{(r^2+g^2)^{3/2}}
		$
		&
		$\displaystyle
		r^2+a_\bullet^2
		-\frac{2\mu}{c^2}
		\frac{r^4}{(r^2+g^2)^{3/2}}
		$
		\\[2ex]
		\hline
		
		\rule{0pt}{6ex}
		Hayward \citep{2016PhRvD..93j4004A}
		&
		$\displaystyle
		r^2
		$
		&
		$\displaystyle
		\frac{
			2(\mu/c^2)r^4
		}{
			r^3+2(\mu/c^2)l_H^2
		}
		$
		&
		$\displaystyle
		r^2+a_\bullet^2
		-\frac{
			2(\mu/c^2)r^4
		}{
			r^3+2(\mu/c^2)l_H^2
		}
		$
		\\[2ex]
		\hline
		
		\rule{0pt}{7ex}
		Sen \citep{1992PhRvL..69.1006S}
		&
		$\displaystyle
		\begin{aligned}
			&r(r+r_0),
			\\
			&r_0 \equiv
			\frac{q_m^2}
			{4\pi\epsilon_0M_\bullet c^2}
		\end{aligned}
		$
		&
		$\displaystyle
		\frac{2\mu r}{c^2}
		$
		&
		$\displaystyle
		r(r+r_0)+a_\bullet^2
		-\frac{2\mu r}{c^2}
		$
		\\[4ex]
		\hline
		
	\end{tabular}
\end{table*}

\section{Hamiltonian and precession corrections to the Schwarzschild case derived using the Lie-transform method}

In Table~\ref{tab:hamiltonian_precession}, we summarise the secular Hamiltonian and corresponding precession corrections derived using the Lie-transform method and subsequently used to construct Figs.~\ref{fig:deviations_non_rotating} and~\ref{fig:deviations_rotating}. We also provide references to previous works that obtained the same precession expressions. 
\begin{table*}[t]
	\centering
	\caption{Summary of the secular Hamiltonian corrections and corresponding precession corrections relative to the Schwarzschild case, together with reference checks for the precession corrections of the spacetimes considered in this work. Here, \(\bar{p}=c^2a(1-e^2)/\mu\) and \(\Delta\psi = \Delta\omega + \Delta\Omega\).}
	\label{tab:hamiltonian_precession}
	\footnotesize
	\setlength{\tabcolsep}{4pt}
	
	\begin{tabular}{
			|p{3.0cm}|
			p{7.1cm}|
			p{4.2cm}|
			p{3.0cm}|
		}
		\hline
		
		\rule{0pt}{4ex}\textbf{Spacetime} &
		\textbf{Secular Hamiltonian correction} &
		\textbf{Precession correction} &
		\textbf{Reference check}
		\\[1.5ex]
		\hline
		
		
		Reissner--Nordstr\"om
		\citep{1916AnP...355..106R,1918KNAB...20.1238N}
		&
		\rule{0pt}{6ex}
		$\displaystyle
		\langle\mathcal H_{\rm RN}\rangle
		=
		\frac{m\mu^2\bar q^2}
		{2a^2c^2\sqrt{1-e^2}}
		$
		&
		$\displaystyle
		\Delta\omega_{\rm RN}
		=
		-\frac{\pi\bar q^2}{\bar p}
		$
		&
		e.g.~\citet{2019arXiv190303432Y}
		\\[1.5ex]
		\hline
		
		
		\rule{0pt}{6ex}
		Bardeen
		\citep{1968qtr..conf...87B}
		&
		$\displaystyle
		\langle\mathcal H_{\rm B}\rangle
		=
		\frac{3\bar g^2m\mu^3}
		{2a^3c^4(1-e^2)^{3/2}}
		$
		&
		$\displaystyle
		\Delta\omega_{\rm B}
		=
		-\frac{9\pi\bar g^2}{\bar p^2}
		$
		&
		e.g.~\citet{2023GReGr..55..114L}
		\\[3ex]
		\hline
		
		
		\rule{0pt}{6ex}
		Hayward
		\citep{2006PhRvL..96c1103H}
		&
		$\displaystyle
		\langle\mathcal H_{\rm H}\rangle
		=
		\frac{\bar l^2m\mu^4(2+e^2)}
		{a^4c^6(1-e^2)^{5/2}}
		$
		&
		$\displaystyle
		\Delta\omega_{\rm H}
		=
		-\frac{6\pi\bar l^2(4+e^2)}{\bar p^3}
		$
		&
		e.g.~\citet{2025JCAP...09..051H}
		\\[3ex]
		\hline
		
		
		\rule{0pt}{6ex}
		Simpson--Visser
		\citep{2019JCAP...02..042S}
		&
		$\displaystyle
		\langle\mathcal H_{\rm SV}\rangle
		=
		-\frac{e^2\bar\lambda^2m\mu^3}
		{4a^3c^4(1-e^2)^{3/2}}
		$
		&
		$\displaystyle
		\Delta\omega_{\rm SV}
		=
		\frac{\pi\bar\lambda^2(2+e^2)}
		{2\bar p^2}
		$
		&
		e.g.~\citet{Zhou2020PrecessingAP}
		\\[3ex]
		\hline
		
		
		\rule{0pt}{6ex}
		Frolov \((\bar l=0.4)\)
		\citep{2016PhRvD..94j4056F}
		&
		$\displaystyle
		\langle\mathcal H_{\rm F}\rangle
		=
		\frac{m\mu^2\bar q^2}
		{2a^2c^2\sqrt{1-e^2}}
		+
		\frac{\bar l^2m\mu^4(2+e^2)}
		{a^4c^6(1-e^2)^{5/2}}
		$
		&
		$\displaystyle
		\Delta\omega_{\rm F}
		=
		-\frac{\pi\bar q^2}{\bar p}
		-\frac{6\pi\bar l^2(4+e^2)}{\bar p^3}
		$
		&
		This work.\textsuperscript{a}
		\\[3ex]
		\hline
		
		
		\rule{0pt}{6ex}
		KS
		\citep{1994NuPhB.429..153K}
		&
		$\displaystyle
		\langle\mathcal H_{\rm KS}\rangle
		=
		-\frac{\bar\alpha^2m\mu^2}
		{4a^2c^2\sqrt{1-e^2}}
		$
		&
		$\displaystyle
		\Delta\omega_{\rm KS}
		=
		\frac{\pi\bar\alpha^2}{2\bar p}
		$
		&
		This work.\textsuperscript{b}
		\\[3ex]
		\hline
		
		
		\rule{0pt}{6ex}
		EMd-1
		\citep{Gibbons:1987ps,Garfinkle:1990qj,1992PhRvD..46.5278K,1992PhRvL..69.1006S}
		&
		$\displaystyle
		\langle\mathcal H_{{\rm EMd}-1}\rangle
		=
		\frac{m\mu^2\bar q^2}
		{2a^2c^2\sqrt{1-e^2}}
		$
		&
		$\displaystyle
		\Delta\omega_{{\rm EMd}-1}
		=
		-\frac{\pi\bar q^2}{\bar p}
		$
		&
		e.g.~\citet{2024EPJC...84.1292H}
		\\[3ex]
		\hline
		
		
		\rule{0pt}{6ex}
		Kerr
		\citep{1963PhRvL..11..237K,1967JMP.....8..265B}
		&
		$\displaystyle
		\langle\mathcal H_{\rm K}\rangle
		=
		\frac{2\chi m\mu^{5/2}}
		{a^{5/2}c^3(1-e^2)}
		$
		&
		$\displaystyle
		\Delta\psi_{\rm K}
		=
		-\frac{8\pi\chi}{\bar p^{3/2}}
		$
		&
		e.g.~\citet{2010PhRvD..81f2002M}
		\\[3ex]
		\hline
		
		
		\rule{0pt}{6ex}
		Kerr--Newman
		\citep{1965JMP.....6..918N}
		&
		$\displaystyle
		\langle\mathcal H_{\rm KN}\rangle
		=
		\frac{m\mu^2\bar q^2}
		{2a^2c^2\sqrt{1-e^2}}
		+
		\frac{2\chi m\mu^{5/2}}
		{a^{5/2}c^3(1-e^2)}
		$
		&
		$\displaystyle
		\Delta\psi_{\rm KN}
		=
		-\frac{\pi\bar q^2}{\bar p}
		-\frac{8\pi\chi}{\bar p^{3/2}}
		$
		&
		e.g.~\citet{2024EPJC...84.1292H}
		\\[3ex]
		\hline
		
		
		\rule{0pt}{6ex}
		(Rotating) Bardeen
		\citep{2016PhRvD..93j4004A}
		&
		$\displaystyle
		\langle\mathcal H_{\rm B}\rangle
		=
		\frac{2\chi m\mu^{5/2}}
		{a^{5/2}c^3(1-e^2)}
		+
		\frac{3\bar g^2m\mu^3}
		{2a^3c^4(1-e^2)^{3/2}}
		$
		&
		$\displaystyle
		\Delta\psi_{\rm B}
		=
		-\frac{9\pi\bar g^2}{\bar p^2}
		-\frac{8\pi\chi}{\bar p^{3/2}}
		$
		&
		e.g.~\citet{2024EPJC...84.1292H}
		\\[3ex]
		\hline
		
		
		\rule{0pt}{6ex}
		(Rotating) Hayward
		\citep{2016PhRvD..93j4004A}
		&
		$\displaystyle
		\langle\mathcal H_{\rm H}\rangle
		=
		\frac{2\chi m\mu^{5/2}}
		{a^{5/2}c^3(1-e^2)}
		+
		\frac{\bar l^2m\mu^4(2+e^2)}
		{a^4c^6(1-e^2)^{5/2}}
		$
		&
		$\displaystyle
		\Delta\psi_{\rm H}
		=
		-\frac{6\pi\bar l^2(4+e^2)}{\bar p^3}
		-\frac{8\pi\chi}{\bar p^{3/2}}
		$
		&
		e.g.~\citet{2024EPJC...84.1292H}\textsuperscript{c}
		\\[3ex]
		\hline
		
		
		\rule{0pt}{6ex}
		Sen
		\citep{1992PhRvL..69.1006S}
		&
		$\displaystyle
		\langle\mathcal H_{\rm Sen}\rangle
		=
		\frac{m\mu^2\bar q_{\rm m}^2}
		{2a^2c^2\sqrt{1-e^2}}
		+
		\frac{2\chi m\mu^{5/2}}
		{a^{5/2}c^3(1-e^2)}
		$
		&
		$\displaystyle
		\Delta\psi_{\rm Sen}
		=
		-\frac{\pi\bar q_{\rm m}^2}{\bar p}
		-\frac{8\pi\chi}{\bar p^{3/2}}
		$
		&
		e.g.~\citet{2024EPJC...84.1292H}
		\\[3ex]
		\hline
		
	\end{tabular}
	
	\vspace{0.5ex}
	\begin{minipage}{0.98\textwidth}
		\footnotesize
		\justifying
		\textsuperscript{a}\,
		To our knowledge, a post-Newtonian treatment of this spacetime for deriving the apsidal precession is not available in the literature; the only analysis we found treats it numerically \citep{BOUZENADA2025116928}. Asymptotically, this spacetime describes a charged black hole with a regularisation parameter \citep{2016PhRvD..94j4056F}. In particular, when the charge vanishes, the Frolov spacetime reduces to the Hayward spacetime, whereas when the regularisation parameter vanishes, it reduces to the Reissner--Nordstr\"om spacetime \citep{2025EPJC...85..848D}. Hence, the leading precession corrections are expected to correspond to those of the Reissner--Nordstr\"om and Hayward cases, as we indeed find.
	\end{minipage}
	
	\begin{minipage}{0.98\textwidth}
		\footnotesize
		\justifying
		\textsuperscript{b}\,
		We are not aware of any previous post-Newtonian treatment deriving the precession for this spacetime. However, the weak-field expansion of the metric resembles that of the Reissner--Nordstr\"om spacetime, with the replacement \(\bar q^2\rightarrow-\bar\alpha^2/2\) \citep[cf.][]{2019arXiv190303432Y,2021EPJC...81..627L}. We therefore expect the apsidal precession correction to take the same form as in the Reissner--Nordstr\"om case under this replacement, as we indeed find.
	\end{minipage}
	
	\begin{minipage}{0.98\textwidth}
		\footnotesize
		\justifying
		\textsuperscript{c}\,
		In \citet{2024EPJC...84.1292H}, we identify a typographical sign error in the expression for the apsidal precession of the rotating Hayward spacetime. Using the conventions of that work, the rotating Hayward result should reduce to the Kerr expression [Eq.~(B1)] when the Hayward deformation parameter vanishes. However, in this limit, the \(p^{-3}\) contribution in Eq.~(B12) differs from the corresponding Kerr term by an overall minus sign. Correcting this sign restores the Kerr limit and yields the same precession expression as derived here.
	\end{minipage}
	
\end{table*}

\end{document}